\documentclass[12pt]{revtex4-1}

\usepackage{amsmath, amsfonts, graphicx, braket}

\begin{document}

\title{ Real--Time Transport in Open Quantum Systems  From $\mathcal{PT}$--Symmetric Quantum Mechanics}
\author{Justin E. Elenewski and Hanning Chen$^{*}$}
\affiliation{Department of Chemistry, The George Washington University, Washington, DC 20052.}
\email{chenhanning@gwu.edu }

\begin{abstract}
Nanoscale electronic transport is of intense technological interest, with applications ranging from semiconducting devices and molecular junctions to charge migration in biological systems. Most explicit theoretical approaches treat transport using a combination of density functional theory (DFT) and non--equilibrium Green's functions. This is a static formalism, with dynamic response properties accommodated only through complicated extensions. To circumvent this limitation, the carrier density may be propagated using real--time time--dependent DFT (RT--TDDFT), with boundary conditions corresponding to an open quantum system. 
Complex absorbing potentials can emulate outgoing particles at the simulation boundary, although these do not account for introduction of charge density.  It is demonstrated that the desired positive particle flux is afforded by a class of $\mathcal{PT}$--symmetric generating potentials that are characterized by anisotropic transmission resonances.  These potentials add density every time a particle traverses the cell boundary, and may be used to engineer a continuous pulse train for incident packets.  This is a first step toward developing a complete transport formalism unique to RT--TDDFT.
\end{abstract}

\maketitle

\section{Introduction}

\par The conjunction of density functional theory (DFT) \cite{Kohn1965} and the non--equilibrium Green's functions (NEGF) method \cite{DiVentra2008, Xue2001} has afforded a tool of unprecedented utility for the computational description of electrical transport in nanoscale devices.  Fruitful applications have extended from metallic and semiconducting constrictions to molecular junctions, with transmission regions spanning a broad swath of chemical parameter space.  While inherently a single--particle approach, many--body corrections may be  phenomenologically included through the DFT+$U$ method or through direct modification of self--energy terms appearing in the NEGF expansion \cite{Timoshevskii2014}.  The scope of these extensions suggests a universal framework for the  atomic--resolution simulation of transport in technologically--relevant materials, which is necessary for the engineering of functional nanoelectronic components. The flexibility and simplicity of this method nonetheless comes at a price, as the calculated conductances are generally between one to two orders of magnitude greater than those observed experimentally \cite{Lindsay2007}.  Furthermore NEGF+DFT calculations employ static, ground--state electronic structure calculations by construction, and hence there is no possibility of calculating time--dependent response properties within this framework.

\par In a first--order attempt to circumvent this limitation, the NEGF method has been expanded to include time--dependent density functional theory (TDDFT) \cite{Runge1984, Stefanucci2004a, Kurth2004}.  While this method is sufficient for model Hamiltonians, self--consistent calculations are difficult to execute \cite{Ke2010}, and self--consistency is requisite for the study of real materials.  One appealing alternative to the NEGF+TDDFT method entails direct propagation of the electronic wavefunction with real--time TDDFT (RT--TDDFT) \cite{Varga2011}.  This scheme likewise ameliorates the cost of NEGF+TDDFT calculations as the numerically expensive determination of Green's functions in the lead regions is no longer necessary \cite{Driscoll2008}.  Nonetheless, a known difficulty associated with RT--TDDFT propagation is the treatment of boundary conditions at the edge of the simulation cell, which must correspond to those of an open quantum system.  Recent investigations with both  NEGF+TDDFT \cite{Driscoll2008, Varga2009, Zhang2013} and RT--TDDFT propagation \cite{Varga2011} have employed a complex absorbing potential in this region to attenuate the wavefunction and avoid spurious reflections.   While previously proposed for transport problems in model systems \cite{Ferry1999, Zhang2007}, these investigations comprise the first application to a realistic case.  The complex potential is itself a non--Hermitian extension to the Hamiltonian that diminishes the net electron density in the system as a function of simulation time.   Reducing the number of electrons within the leads will alter the contribution from Hartree and exchange--correlation terms in the DFT Hamiltonian, and lead to a jamming process in which transport no longer occurs as the system becomes ionized.  Thus, the absorbing potentials only comprise  half of the framework required for a comprehensive treatment of transport, as generating potentials for incoming charge carriers are also required.

\par The addition of a complex potential $\hat{V}_\text{cplx}$ to a quantum system has an unusual effect on the time evolution of a state vector.  Consider a non--Hermitian Hamiltonian $\hat{H} = \hat{H}_0 + \hat{V}_\text{cplx}$ in which $\hat{V}_\text{cplx}$ may be arbitrarily applied, and let $\hat{H}_0$ be a Hermitian Hamiltonian which is applicable at all times.  Furthermore, let $\ket{\psi(x,t = 0)}$ be an initial eigenstate of $\hat{H}_0$ when $\hat{V}_\text{cplx}$ is zero.  As $\ket{\psi(x,t)}$ propagates, assume that a purely imaginary $\hat{V}_\text{cplx} \approx i\Gamma \neq 0$ is turned on starting at  time $t_1$ and turned off at time $t_2 > t_1$.  In the course of this process, the state vector evolution is afforded by the operator $\hat{U} (t', t) = \exp[-i\hat{H}(t' - t)/\hbar]$ so that 
$\ket{\psi(x,t_2)} = \hat{U}(t_2, t_1) \ket{\psi(x,t_1)}$, or explicitly

\begin{eqnarray}
\ket{\psi(x,t_2)} &=& \exp[-i(\hat{H_0} + i\Gamma) (t_2 - t_1) / \hbar] \ket{\psi(x,t_1)}\\
&=& \exp[-i\hat{E} (t_2 - t_1) / \hbar] \exp[\Gamma (t_2 - t_1) / \hbar] \ket{\psi(x,t_1)}.
\end{eqnarray}

\noindent The first term in the product is simply the time evolution operator for the system under the action of $\hat{H}_0$ alone, whereas the second term characterizes the effect of the complex potential.  Taking the inner product $\braket{\psi(x,t_2) \vert \psi(x,t_2)} = \exp[2\Gamma(t_2 - t_1) / \hbar] \braket{\psi(x,t_1) \vert \psi(x,t_1)}$, it is clear that the norm of the particle is rescaled by a factor of $\exp[2\Gamma(t_2 - t_1) / \hbar]$.  Furthermore, just as the Hamiltonian is no longer Hermitian in the presence of $\hat{V}_\text{cplx}$, the evolution operator $\hat{U}(t',t)$ ceases to be unitary.

\par If the complex potential strength $\Gamma < 0$, the norm of the state vector is decreased and  the effective particle number in the system is diminished.  This behavior is key when using the complex potential to mimic open boundary conditions that accommodate an incoming or outgoing particle flux \cite{Varga2007, Paul2007, Berggren2010, Wibking2012, Fortanier2014, Wahlstrand2014, Zhu2014} as well as for the treatment of resonances in wavepacket propagation  \cite{Moiseyev1998, Muga2004, Moiseyev2011}, and atomic \cite{Sahoo2000}, and nuclear systems \cite{Masui2002}.  Conversely, if the sign of the potential is flipped so that $\Gamma > 0$,  the potential then generates norm for a given state, which may be conceptualized  as the addition of particles to the system.  The presence of such `source' and ``sink' terms is a general property of  simple non--Hermitian extensions \cite{Ferry1999, Berggren2010, Wahlstrand2014}.  One particularly useful category of these theories are the $\mathcal{PT}$--symmetric Hamiltonians, in which the condition of Hermiticity is relaxed in favor of symmetry under conjugation by the product of the parity $\hat{\mathcal{P}}$ and time reversal $\hat{\mathcal{T}}$ operators \cite{Bender1998, Bender2002, Bender2007}.  The construction of $\mathcal{PT}$--symmetric theories extends the spectrum of effective Hamiltonians which may be employed to describe open quantum systems, and has led to several experimentally verified predictions in optics \cite{Guo2009,Ruter2010, Sun2014}.

\par Given these considerations, it is natural to ask if an appropriately constructed $\mathcal{PT}$--symmetric theory can fully mimic the progressive, time--dependent addition and removal of particles in an open system, particularly in a manner that does not require continuous tuning of source and sink terms.  By constructing an appropriate scheme using recent lessons from $\mathcal{PT}$--symmetric optics this question is answered in the affirmative.  In particular, it is demonstrated that particle generation and attenuation may be integrated under conditions corresponding to either an applied voltage or current bias.  Analytical and numerical results are evaluated for the particular problem of real--time one--dimensional wavepacket propagation to demonstrate the efficacy of this method.  It is further demonstrated that such a framework may be naturally extended to density functional theory  without  limitations.

\section{Analytical Considerations}

\subsection{$\mathcal{PT}$--Symmetric Quantum Mechanics}

\par In a $\mathcal{PT}$--symmetric quantum system, the requirement that the Hamiltonian be Hermitian is relaxed to a more general conjugation condition \cite{Bender2002, Bender2007}.  Specifically, a new operator $\hat{\mathcal{P}}\hat{\mathcal{T}}$ is introduced as a product of the  parity  $\hat{\mathcal{P}}: \{\hat{p}, \hat{x}, i\mathbb{Id}\} \mapsto \{-\hat{p}, -\hat{x}, i\mathbb{Id}\}$ and  time--reversal $\hat{\mathcal{T}}: \{\hat{p}, \hat{x}, i\mathbb{Id}\} \mapsto \{-\hat{p}, \hat{x}, -i\mathbb{Id}\}$ operations, such that the composite operator $\hat{\mathcal{P}}\hat{\mathcal{T}}$ and the new Hamiltonian $\hat{H}_{PT}$ share a common set of eigenfunctions.   Invariance under Hermitian conjugation is replaced with the commutator $[\hat{\mathcal{P}}\hat{\mathcal{T}},\hat{H}_{PT}] = 0$.  Note that $\hat{H}_{PT}$ need not commute with the action of $\hat{\mathcal{P}}$ and $\hat{\mathcal{T}}$ alone, but only with the operator product.  When these conditions are collectively satisfied, the system is said to possess exact or unbroken $\mathcal{PT}$--symmetry \cite{Bender1998,Bender2002}.    If the potential is only a function of the particle position, the Hamiltonian $\hat{H}_{PT}$ may be written in the elementary form $\hat{H}_{PT} = \hat{p}^2 / 2m + \hat{V}_{PT}(\hat{x})$, whereupon $\mathcal{PT}$--symmetry requires that $\hat{V}_{PT}(\hat{x}) = \hat{V}_{PT}^* (-\hat{x})$ with the asterisk denoting complex conjugation.  Accordingly, the potential may be expanded as $\hat{V}_{PT}(\hat{x}) = \text{Re}[\hat{V}_{PT}(\hat{x})] + i\text{Im}[\hat{V}_{PT}(\hat{x})]$, where the real and imaginary parts are even and odd functions of $\hat{x}$, respectively.
\par Despite the presence of an imaginary potential, the unbroken symmetry phase of a $\mathcal{PT}$--symmetric theory is characterized by a real eigenvalue spectrum.  Conversely, in the so--called broken symmetry phase, $\hat{\mathcal{P}}\hat{\mathcal{T}}$ and $\hat{H}_{PT}$ cease to share a common eigenfunction space and the roots of the eigenvalue problem become complex.  This spectral behavior  has been systematically investigated for several potentials, including those of the form $V(x) = \alpha x^2(ix)^\nu$   with $\alpha \in \mathbb{R}$ and $\nu \in \mathbb{N}$ \cite{Bender1998, Bender2012}.  It is conjectured that an arbitrary $\mathcal{PT}$--symmetric complex potential $V(x)$ must be analytic to possess a real spectrum \cite{Dorey2001, Bender2008b}, though other more stringent requirements may also apply \cite{Bender2007}.  This surprising observation of a well--defined real eigenvalue spectrum led to the proposition that $\mathcal{PT}$ symmetry could represent a generalization of quantum mechanics, especially when formulated in terms of an inner product structure with additional symmetries \cite{Bender2002}.  

\par Nonetheless, a local $\mathcal{PT}$ symmetry allows arbitrarily fast quantum state evolution \cite{Bender2007}, including superluminal propagation \cite{Lee2014}, thus limiting the applicability of such Hamiltonians as a fundamental extension of quantum mechanics.  Furthermore, global $\mathcal{PT}$ symmetric Hamiltonians are isomorphic to conventional Hermitian Hamiltonians for finite--dimensional systems \cite{Mostafazadeh2002, Mostafazadeh2002b, Mostafazadeh2002c, Mostafazadeh2003, Mostafazadeh2007}, differing only in their unconventional definition of the inner product.  In spite of such restrictions, these structures afford a mathematically useful framework for effective theories in the condensed matter realm, particularly for open quantum systems \cite{ Varga2007, Paul2007, Rotter2009, Berggren2010,Wibking2012, Fortanier2014, Wahlstrand2014, Zhu2014}, for the computational treatment of resonances in wavepacket dynamics and scattering \cite{Moiseyev1998, Muga2004, Moiseyev2011} and for light propagation in certain optical lattices \cite{Musslimani2008, Mostafazadeh2009, Ramezani2010, Lin2011}.  Several experimental realizations of $\mathcal{PT}$--symmetry have been explored in this optical context, including loss--induced optical transparency \cite{Guo2009} and left--right asymmetric power oscillations \cite{Ruter2010} in a nonlinear optical device, unidirectional invisibility in a $\mathcal{PT}$--symmetric optical lattice, and the existence of coherent perfect absorbers assembled using passive optical components \cite{Sun2014}.

\subsection{Quantum Transport in $\mathcal{PT}$--Symmetric Potentials}

\par A characteristic of  $\mathcal{PT}$--symmetric non--Hermitian theories is the presence of `source' and `sink' terms for the wavefunction norm \cite{Ferry1999,Berggren2010,  Wahlstrand2014}.  While a self--consistent norm has been devised for $\mathcal{PT}$--symmetry \cite{Bender2002}, we are interested in physical systems for which $\hat{H}_{PT}$ is an effective Hamiltonian and thus do not adopt this definition.  Accordingly, denote by $\mathcal{N} = \braket{\psi(x,t) \vert \psi(x,t)}$ the $\mathbb{L}^2 (\mathbb{R})$ norm of $\psi(x,t)$ in a Hilbert space $\mathcal{H}$.  Calculating the time dependence directly in terms of the $\mathcal{PT}$--symmetric potential $V_{PT}(x) = \text{Re}[V_{PT}(x)] +  i\text{Im}[V_{PT}(x)]$ yields

\begin{eqnarray} \label{attenrate}
\frac{d\mathcal{N}(t)}{dt} &=& \int_{-\infty}^\infty dx\, \frac{d}{dt} \left(\psi^* (x,t) \psi(x,t)\right)) \\
&=& \int_{-\infty}^\infty dx\, \left(\frac{d\psi^*(x,t)}{dt} \psi(x,t) + \psi^*(x) \frac{d\psi(x,t)}{dt}\right) \\
&=& \frac{1}{i\hbar} \int_{-\infty}^\infty dx\,\left(\psi^*(x,t) \hat{H}_{PT} \psi(x,t) - \hat{H}_{PT}^\dagger \psi^*(x) \psi(x)\right) \\
&=& \frac{1}{\hbar} \braket{\psi\vert(\hat{H}_{PT} - \hat{H}_{PT}^\dagger)\vert\psi} \\
&=& \frac{2}{\hbar} \braket{\psi \vert \text{Im}(\hat{V}_{PT}) \vert \psi}
\end{eqnarray}

\noindent The condition for norm attenuation,  $d\mathcal{N}(t)/dt < 0$,  requires that $\braket{\psi \vert \text{Im}(\hat{V}_{PT}) \vert \psi} < 0$.  A similar condition for norm generation applies when $d\mathcal{N}(t)/dt > 0$, indicating that the imaginary part alone determines the `source' or `sink' behavior.   Interestingly, since this process is contingent upon an expectation value, the source term is incapable of generating norm in the absence of some finite probability amplitude within the spatial extent of the potential.  Once a state is completely attenuated, it many never be recovered by a generating term.  

\par  This relation has an important association with transport properties. Specifically, the net outgoing change in norm for all particles in a many--particle system $\partial \mathcal{N}_T / \partial t = \sum_i \partial \mathcal{N}_i / \partial t$ due to the presence of the imaginary potential must be equal to the integrated divergence of the current  through the system  $-\int_V \nabla \cdot \vec{j}(\vec{x},t) dV = \partial \mathcal{N}_T(t) / \partial t$, where $\vec{j}(\vec{x},t)$  is the local probability current density

\begin{equation}
\vec{j}(\vec{x},t) = \frac{e\hbar}{2mi} \sum_{i} \left[\psi_i^* (\vec{x},t) \nabla \psi_i(\vec{x},t) - \psi_i (\vec{x},t) \nabla \psi_i^*(\vec{x},t)\right].
\end{equation}

\noindent This relationship between norm and net current is obtained by directly integrating the continuity equation for the particle density.

\subsection{Complex Absorbing Potentials}

\par Complex absorbing potentials have been systematically developed using functional forms including  linear and step potentials \cite{Neuhasuer1989}, higher--order polynomials \cite{Vibok1992, Riss1996, Ge1997, Poirier2003, Poirier2003b}, exponential \cite{Vibok1992, Vibok1992b} and hyperbolic functions \cite{Kosloff1986}, as well as through functions with singular behavior at isolated points in the complex plane \cite{Brouard1994, Manolopoulos2002}.  These investigations do not suggest a universal `optimal absorbing potential,' however, the criteria necessary for an effective absorber may be distinguished. In particular, complex polynomial potentials significantly enhance absorption over purely imaginary polynomial terms, particularly in low energy cases where the deBroglie wavelength of the incident wavepacket is comparable to the characteristic length of the absorbing region \cite{Ge1997}.  Adding a negative real component to the potential will increase the energy of the incident particle and thereby reduce the wavelength of the packet,  enhancing norm attenuation by the imaginary part while concurrently reducing reflection.  It should be noted that the Wentzel--Kramers--Jeffreys--Brillouin (WKJB) approximation yields quantitatively inaccurate results where $\lambda / L \geq 1$, which is the domain of interest for most applications of absorbing potentials \cite{Ge1997}. Despite these limitations, potentials optimized in the semiclassical limit will be utilized as--is, with the assumption that general trends in absorbing efficiency are transferrable.  This approximation is found to be sufficient, provided that the numerical parameters defining the potential are adjusted at runtime.

\par
\begin{figure} 
\centering
 \includegraphics[scale=1.0]{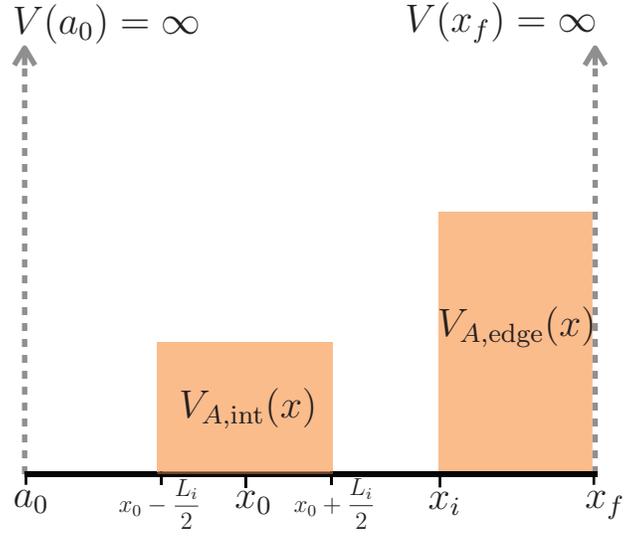} 
\caption{Contexts for the use of complex absorbing potentials in an infinite square well.  The potential $V_{A,\text{edge}}(x)$ is employed to absorb wavepackets impinging on the simulation cell boundary to mimic the effect of a particle leaving an open system.   Conversely, the  $V_{A,\text{int}}(x)$ absorbs wavefunction norm incident from either side of the potential, with a net effect of  ``disconnecting'' two regions of the simulation.}
\label{absorb_schematic}
\end{figure}

\begin{figure} 
\centering
 \includegraphics[scale=1.0]{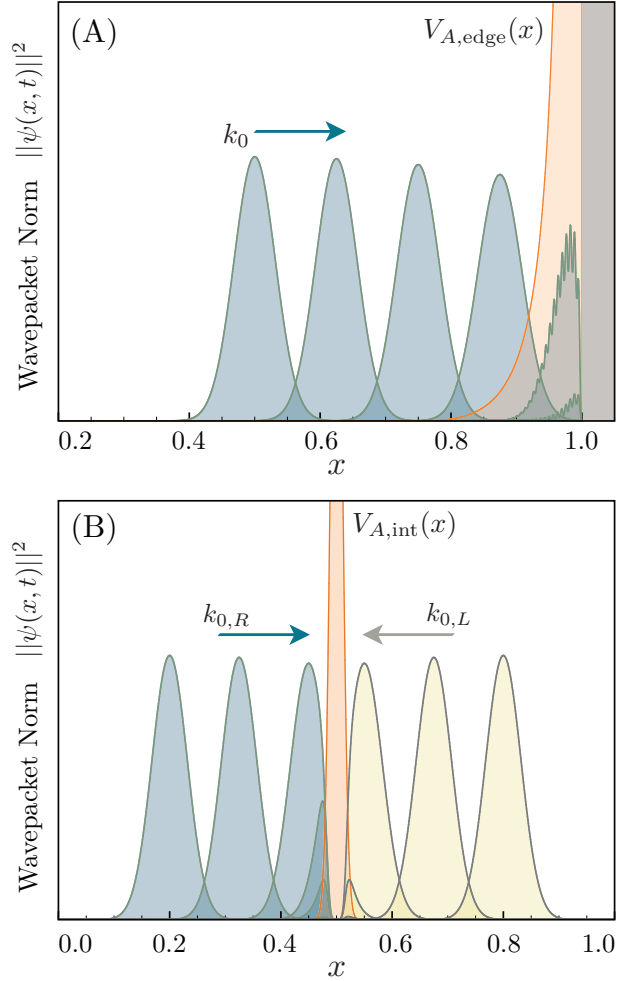} 
\caption{Wavepacket attenuation for two distinct classes of complex potentials, as demonstrated through numerical wavepacket propagation.  (A) Propagation of a Gaussian wavepacket with  $k_0 = 500$, $\sigma^2 = 0.001$, and $x_0 = 0.5$ (blue) into a potential $V_{A,\text{edge}}(x)$ (orange) with singularity at the cell boundary.  The potential switches on at $x = 0.75$ with a width $L = 0.25$ and a strength of $E_\text{min} = 4.0$.  Each packet envelope corresponds to a configuration advanced by $\Delta t = 2.5 \times 10^4$ units.  (B) Propagation of a right--moving Gaussian wavepacket $k_{0,R} = 500$ and $x_{0,R} = 0.2$ (blue) alongside a left--moving Gaussian wavepacket $k_{0,L} = -k_{0,R}$ and $x_{0,R} = 0.8$ (yellow)  into a Gaussian absorbing potential $V_{A,\text{int}}(x)$ (orange).  The potential is applied for all $x \in [0.4,0.6]$ with a width $\alpha^2 = 1.0 \times 10^{-4}$ and strength $V_0 = 5.0 \times 10^3$.  Each packet envelope corresponds to a configuration advanced by $\Delta t = 2.5 \times 10^4$ units, with an additional configuration shown at $t = 1.35 \times 10^5$ units for the rightmoving packet.  Packets are propagated using default propagation parameters. }
\label{absorb_sim}
\end{figure}

 A further consideration is related to the specific application of a given potential within a simulation. In the first case, a complex potential may be located at the boundary of the simulation cell to absorb particles leaving the system [Fig. \ref{absorb_schematic}].  Such a potential should switch on smoothly outside of the interaction region and attain larger values as the distance from this region increases.  A smooth profile is essential to minimize reflections, as any discontinuous step will be reflection generating \cite{Poirier2003, Poirier2003b}.  While satisfied by simple cases such as complex polynomials, a particularly efficacious attenuator is the potential $V_{A, \text{edge}} (w) = -i E_{\text{min}} f(w)$, with 

\begin{equation}
f(w) = \left(1 - \frac{16}{c^3}\right)w - \frac{1}{c^2}\left(1 - \frac{17}{c^3}\right)w^3 + 4\left(\frac{1}{(c-w)^2} - \frac{1}{(c+w)^2}\right)
\end{equation}

\noindent where the variable $w = 2 \delta k_\text{min} (x-x_i)$ has been introduced.   In this case, $x_i$ marks the incoming boundary of the potential, and $x_f$ corresponds to the edge of the simulation cell, $E_\text{min}$ is the lowest energy of interest for an incident particle, and $\delta k_\text{min} = \sqrt{2 m E_\text{min} / \hbar^2}$ is the corresponding wavevector.  In order that the potential become singular as $x \longrightarrow x_f$, it is necessary to set $\delta k_\text{min} = c / 2L$, where $L = (x_f - x_i)$ is the length of the potential \cite{Manolopoulos2002}. This particular functional form was constructed as the solution of a semiclassical differential equation derived for plane wave scattering from a complex potential.  The sum of reflection and transmission coefficients $\vert R\vert^2 + \vert T \vert^2$ was minimized as a constraint during construction, thereby ensuring optimal absorption.  Furthermore, the divergent growth as $x \longrightarrow x_f$ ensures complete attenuation before the cell boundary is reached [Fig. \ref{absorb_sim}(a)].

\par While the aforementioned potential is ideal for boundary attenuation, it may also become necessary to attenuate the wavefunction within the interaction region [Fig. \ref{absorb_schematic}].  Such a potential must be symmetric to ensure isotropic scattering from each side and bidirectionally smooth to minimize reflections.  The simplest such choice is a Gaussian function

\begin{equation}
V_{A,\text{int}}(x) = -i V_0 e^{-(x - x_0)^2 / 2\alpha^2}
\end{equation}

\noindent where $\alpha^2$ delimits the spread of the Gaussian. Since the spatial extent of this potential is infinite, it must be defined on a piecewise subdomain $\vert x - x_0 \vert \leq L/2$, where $\alpha$ and $L$ are chosen so that the Gaussian becomes sufficiently small at $x = x_0 \pm L/2$ [Fig. \ref{absorb_sim}(b)].  Taken together, these two functional forms comprise a sufficient armamentarium of absorbing potentials to handle any scenario that may be encountered in a routine transport simulation.  As a final note, if necessary and irrespective of the application, a negative real part may always be added to decrease the deBroglie wavelength of the incident particle and enhance absorption.

\subsection{$\mathcal{PT}$--Symmetric Generating Potentials}

\par
\begin{figure} 
\centering
 \includegraphics[scale=1.0]{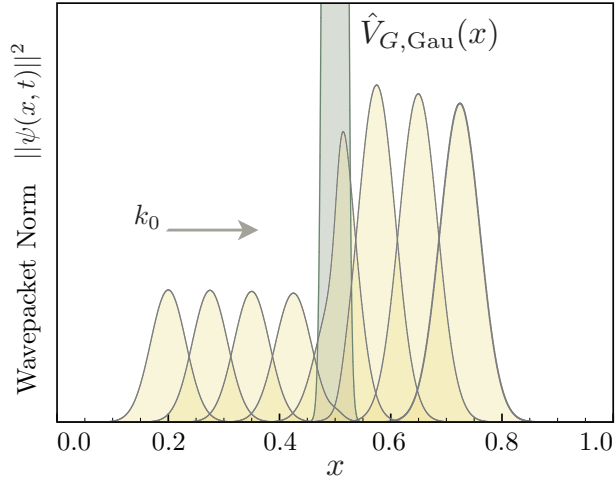} 
\caption{Enhancement of packet norm by a complex generating potential of the form $\hat{V}_{G,\text{Gau}} = iV_0 e^{-(x-x_0)^2 / 2 \alpha}$.  The incident packet has a wavevector $k_0 = 500$ and width $\sigma^2 = 0.001$ with the envelope initially centered at $x_0 = 0.2$, while the potential parameters are $V_0 = 250$ and $\alpha^2 = 1.0 \times 10^{-4}$, with a center at $x_0 = 0.5$.  In the course of propagation, the norm of the packet increases so that $\braket{\psi(x,t_f) \vert \psi(x,t_f)} = 2.72$.  Wavepackets are plotted at time steps ranging from $t_i = 0$ to $t_f = 2.1 \times 10^5$ in units of $\Delta t = 3.0 \times 10^4$ using default  parameters.  }
\label{gaussgrow}
\end{figure}

 Just as a negative imaginary potential attenuates the wavefunction norm, a positive imaginary potential will increase the wavefunction norm.  Consider a simple experiment in which a Gaussian wavepacket of unit norm $\braket{\psi(t_2) \vert \psi(t_2)} = 1$ is incident on a potential of the form 

\begin{equation}
V(x) = \left\{ \begin{array}{ccc} iV_0 e^{-(x-x_0)^2/2\alpha},  & & -L/2 \leq x \leq L/2 \\
               0,                              & & \text{otherwise} \end{array}  \right.
\end{equation} 

\noindent at $t = t_1$, and emerges from the potential later at time $t = t_2$.  During transmission, the norm of the packet will have been increased in magnitude, however, the shape of the packet will be unchanged [Fig. \ref{gaussgrow}].  If an additional unit of norm is added to the packet so that $\braket{\psi(t_2) \vert \psi(t_2)} = 2$, this may be interpreted as the addition of a second particle to the system.  Nonetheless, this scenario is unphysical as the particles coincide spatially and copropagate under time evolution.  To avoid such complications, the use of these potentials in transport calculations requires systematic system and bias dependent tuning \cite{Varga2007, Wibking2012}.

\par A more suitable option is provided through $\mathcal{PT}$--symmetric potentials possessing anisotropic transmission resonances (ATR), also known as `unidirectional invisibility.'  Such potentials were first theoretically investigated in the context of optical heterostuctures and Bragg gratings characterized by alternating gain / loss regions \cite{Lin2011}, with subsequent experimental realization in a temporal optical lattice \cite{Regensburger2012}.  At the spontaneous $\mathcal{PT}$--symmetry breaking point, these systems permit near--perfect transmission of a wave incident from either side while simultaneously reflecting waves at one boundary and being reflectionless at the other.  This anisotropy is a manifestation of the generalized unitarity condition satisfied by $\mathcal{PT}$--symmetric potentials \cite{Ge2012}. Furthermore, the reflecting side of such an optical structure may exhibit enhanced gain; that is, the reflected wave may have an amplitude greater than that of the incident wave.  This phenomenon has direct implications for matter--wave scattering, as the paraxial approximation to the equation of motion for propagation of the electric field $E(x)$ in a medium is formally equivalent to the Schr\"{o}dinger equation.  In the case of an optical heterostructure, the variation in the index of refraction $n$  occurs longitudinal to the incident wave, and this assumes the form of a Helmholtz equation

\begin{equation}
\frac{\partial^2 E(x)}{\partial x^2}  + k^2 \left(\frac{n}{n_0}\right)^2 E(x) = 0
\end{equation}

\noindent where the wavevector $k = n_0 \omega / c$, the index of refraction of the surrounding medium is $n_0$, $c$ is the speed of light in vacuum, and $\omega$ is the angular frequency of the wave. Introducing the convention that $(n / n_0)^2 = (1 + 2 V_\text{ATR}(x))$ establishes a formal connection to the Schr\"{o}dinger equation and the quantum case.  To mimic the aforementioned heterostructures assume that the complex potential, and hence index of refraction, acts over a range $0 \leq x \leq L$ and has the functional form

\begin{eqnarray} \label{ATRpot}
V_\text{ATR}(z) &=& V_A \cos (2\beta x) + i V_B \sin(2\beta x)\\
     &=& V_0 e^{2i\beta x} 
\end{eqnarray}

\noindent where $V_A = V_B = V_0$ is assumed in the second line and  $\beta = \pi / \Lambda$ for a lattice of spatial periodicity $\Lambda$.  It is clear that this potential satisfies the condition $V_\text{ATR}(x) = V_\text{ATR}^*(-x)$ as required for $\mathcal{PT}$--symmetry.  Note that the choice $V_A = V_B$ places the system at the critical point for $\mathcal{PT}$--symmetry breaking, with a real energy spectrum retained for $V_B / V_A \leq 1$.  This ratio likewise controls the left/right--reflection asymmetry.  Within the coupled--mode approximation \cite{Lin2011}, the transmission coefficient is found to vanish for wavevectors with $\delta = \beta - k = 0$, as does the reflection coefficient for left--incident (right--incident) plane waves.  Conversely, the reflection coefficient for right--incident (left--incident) plane waves grows as $L^2 (k V_A)^2$.  Note that, unlike the Schr\"{o}dinger equation, the ``potential'' appearing in the Helmholtz equation is energy dependent through the $k$ terms.  For shallow gratings this dependence is negligible and hence the equivalence is exact \cite{Kulishov2005}.

\par The aforementioned analysis of invisibility is nonetheless performed in an approximate regime.  An exact solution of the  Schr\"{o}dinger equation at the $\mathcal{PT}$--symmetry breaking point

\begin{equation}
\frac{\partial^2 \psi(x,t)}{\partial x^2} + \frac{2m}{\hbar^2} \left(E - \hat{V}_{ATR}(x)\right)\psi(x,t) = 0 
\end{equation}

\noindent with $\hat{V}_{ATR}(x) = V_0 \left[\cos(2\beta x) + i\sin(2\beta x)\right]$ for $0 < x < L$ may be obtained.   Performing a change of variables to $y = (\Lambda \sqrt{V_0} / \pi) \exp[i \pi x / \Lambda]$, the Schr\"{o}dinger equation becomes

\begin{equation}
y^2 \frac{d^2 \psi}{dy^2} + y \frac{d\psi}{dy} - (y^2 + \nu^2)\psi = 0
\end{equation}

\noindent where $\nu = k \Lambda / \pi$, and the convention that $\hbar = 2m = 1$ has been adopted for convenience of notation.  In this case $k$ is the wavevector associated with the momentum of the quantum particle through $p = \hbar k$.  This is a Bessel equation with  solutions $\psi_k (x) = I_\nu (y)$ and $\psi_{-k} (x) = I_{-\nu} (y)$ given in terms of the modified Bessel functions of the first kind.  These functions remain linearly independent provided that $k \Lambda / \pi$ is not an integer \cite{Graefe2011,Longhi2011,Jones2012}.  It is significant to note that these solutions are not orthogonal in the conventional sense, however, they are orthogonal under the $\mathcal{PT}$--symmetric inner product.  At exceptional points where $\nu = n \in \mathbb{N}$, there exists a spectral singularity  \cite{Mostafazadeh2009, Longhi2010}, whereupon $\psi_{\pm k}(x)$ become degenerate.  To resolve this situation, the solutions must be extended by the addition of Jordan associated functions \cite{Graefe2011}.  This general solution will be neglected herein, with the approximation of solutions at all points by  Bessel functions in subsequent calculations. As before, $\beta = \pi/\Lambda$  with the provision that $L=N \Lambda$ with $N \in \mathbb{N}$, corresponding to a $\mathcal{PT}$--symmetric crystal $N$ cells in length. 

\par  Analysis of these solutions indicates that invisibility is not exact  for $\delta = \beta - k \neq 0$, with a nontrivial breakdown of this assumption particularly apparent beyond a critical length $L_c = (2\pi^3 / V_0^2 \Lambda^3)$ \cite{Longhi2011}.  This observation is consistent with the modified unitarity condition for  $\mathcal{PT}$--symmetric potentials, $T - 1 = \pm \sqrt{R_L R_R} $, which bounds the deviation from ideal behavior  \cite{Ge2012}.   Numerical results further indicate that the transmission $T$ and unenhanced reflection coefficients $R_L$ oscillate rapidly as a function of $\delta$, however, the amplitude of this oscillation remains small.  The enhanced reflection coefficient $R_R$, on the other hand, 
 affords a strong enhancement only within a narrow window of values about $\delta = 0$ \cite{Jones2012}.

\par
\begin{figure} 
\centering
 \includegraphics[scale=1.0]{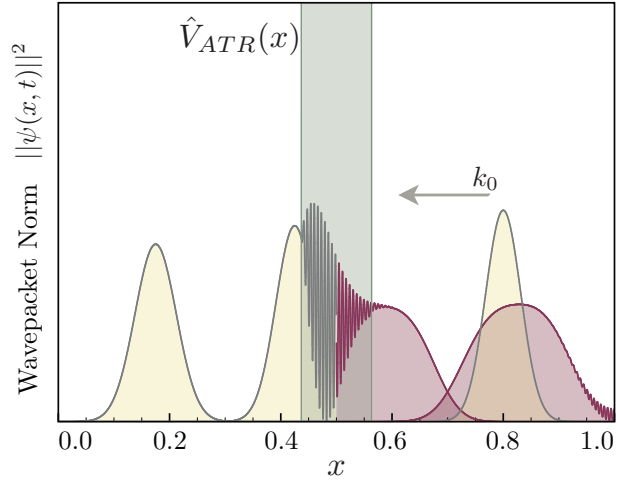} 
\caption{Interaction of a wavepacket with a complex potential $\hat{V}_{ATR}(x)$ exhibiting anisotropic transmission resonances on a scale broader than the incident wavepacket.  Note that the reflected, generated packet (purple) is enhanced in width.  The incident packet (yellow) has a wavevector $k_0 = 500$ and width $\sigma^2 = 0.001$ with the envelope initially centered at $x_0 = 0.2$, while the potential strength is $V_0 = 6500$, the unit cell spacing is $\Lambda = 6.28 \times 10^{-3} = \pi / k$, and the total width is $L = 20 \Lambda$ with a center at $x_0 = 0.5$.   Wavepackets are plotted at time steps of $t_i = 0$, $t = 1.5 \times 10^5$, and $t = 2.5 \times 10^5$.  Wavepacket propagation is performed using default parameters.  }
\label{packet_saturate}
\end{figure}

 A unique phenomenon is observed when a Gaussian wavepacket is incident on an ATR potential [Fig. \ref{packet_saturate}].  Assuming the packet is incident on the generating interface of the ATR, the reflected wave eventually saturates in amplitude, emerging with an extended, flattened peak.  This extrusion process occurs during the entire period for which the maximum of the incident packet remains under the barrier. This phenomena was first observed in numerical simulations and perturbative calculations \cite{Longhi2010} and later rationalized in terms of the Jordan--block structure of the eigenfunction space for the potential \cite{Graefe2011}.  Physically, this saturation occurs due to the presence of spectral singularities, with the resultant spectral broadening causing a saturation in the secular growth of scattered waves.  While a linear scaling behavior would be expected at  this point, the excited Jordan associated functions grow linearly to precisely compensate the decrease in contribution from the nondegenerate states, leading to the stalled growth.   Formally, this corresponds to an incident Gaussian packet being reflected as a sum of error functions \cite{Jones2011} and thus the incident pulse is lengthened into an extended packet of peak width $\sim L$ upon reflection \cite{Jones2012}.

\subsection{Wavepacket Propagation and Transmission / Reflection Coefficients}

\par Consider a barrier penetration problem in which a free particle is incident on an isolated potential $\hat{V}_{PT}$ of width $L$, with the potential permitting anisotropic transmission resonances as per Eq. (\ref{ATRpot}).  In a first order approach,  the  wavefunctions in the left and right regions may be expanded in terms of plane--wave eigenstates

\begin{equation}
\psi(x) = \left\{ 
\begin{array}{lcc} 
\psi_{L,k}(x) = \frac{1}{\sqrt{2\pi}}(A_L e^{ikx} + B_L e^{-ikx}) && x \leq 0 \\
\psi_{R,k}(x) = \frac{1}{\sqrt{2\pi}}(A_R e^{ik(x-L)} + B_R e^{-ik(x-L)})&& x \geq L
\end{array}\right.
\end{equation}

\noindent The wavefunctions on either side of the scattering region are linked through the transfer matrix $\hat{M}(k)$ with components

\begin{equation}
\left( \begin{array}{c} A_R \\ B_R\end{array}\right) = 
\left( \begin{array}{cc} M_{11}(k) & M_{12}(k) \\ M_{21}(k) & M_{22}(k) \end{array}\right) \left( \begin{array}{c} A_L \\ B_L\end{array}\right)
\end{equation}

\noindent from which the transmission amplitude $t_R  = 1 / M_{22}$ as well as left $r_L = -M_{21} / M_{22}$ and right $r_R = M_{12} / M_{22}$ reflection amplitudes are readily obtained.   Evaluating the  Bessel function solutions for this potential at the boundaries, the transfer matrix $M(k)$ is constructed  explicitly \cite{Longhi2011}:

\begin{eqnarray}
M_{11}(k) &=& \cos (kL) + i \frac{\Lambda \sin(kL)}{2k \sin(\pi \nu)} \left(k^2 Q_1 Q_2 - V_0 D_1 D_2\right) \\
M_{12}(k) &=& -i \frac{\Lambda \sin(kL)}{2k \sin(\pi \nu)} \left(V_0 D_1 D_2 + k^2 Q_1 Q_2 + k \sqrt{V_0} \left(D_1 Q_2 + D_2 Q_1\right) \right) \\
M_{21}(k) &=& i \frac{\Lambda \sin(kL)}{2k \sin(\pi \nu)} \left(V_0 D_1 D_2 + k^2 Q_1 Q_2 - k \sqrt{V_0} \left(D_1 Q_2 + D_2 Q_1\right) \right) \\
M_{22}(k) &=& \cos (kL) - i \frac{\Lambda \sin(kL)}{2k \sin(\pi \nu)} \left(k^2 Q_1 Q_2 - V_0 D_1 D_2\right) 
\end{eqnarray}

\noindent where the notation 

\begin{equation}
\begin{array}{cc}
Q_1 = I_\nu (\Delta), & D_1 = \partial_x I_\nu(\Delta) \\
Q_2 = I_{-\nu} (\Delta), & D_2 = \partial_x I_{-\nu}(\Delta)
\end{array}
\end{equation}

\noindent has been introduced with $\Delta = \Lambda \sqrt{V_0} / \pi$ and $2m = 1$.  Similar solutions for other masses may be recovered through the substitution $k \mapsto k / \sqrt{2m}$ and consistent rescaling.

\par These relations afford the reflection and transmission coefficients for plane--wave scattering through the $\mathcal{PT}$--symmetric media.  Nonetheless, the corresponding result for wavepacket transmission will differ substantially, especially when the packet width is comparable to the extent of the $\mathcal{PT}$--symmetric region.  To derive the corresponding coefficients for a finite--width packet, create an initial  envelope of width $\sigma^2$ and wavevector $k_0$ centered at $x=a$:
 
\begin{equation}
\phi(x,0) = \frac{1}{(\pi\sigma^2)^{1/4}} e^{i k_0 (x - a)} e^{-(x-a)^2 / 2\sigma^2},
\end{equation}

\noindent from which the momentum--space representation may be obtained via a Fourier transform

\begin{eqnarray}
\phi(k,0) &=& \frac{1}{\sqrt{2\pi}}\int_{-\infty}^{\infty} \phi(x,0) e^{-ikx} \,dx\\
&=& \left(\frac{\sigma^2}{\pi}\right)^{1/4} e^{(k-k_0)^2\sigma^2 / 2} e^{-ika}.
\end{eqnarray}

\noindent Using this expression, the wavefunction may be synthesized in terms of the plane wave eigenfunctions $\psi_k(x)$

\begin{equation} \label{packeteqn}
\psi(x,t) = \int \psi_k(x) \phi(k,0) e^{-iE(k)t} \, dk,
\end{equation}

\noindent where $E(k) = k^2$ is the energy of a given plane--wave component.  For illustrative purposes, assume that $\psi_k(x) = C(k) (2\pi)^{-1/2}  e^{ikx}$ for a rightmoving packet.  Expanding Eq. (\ref{packeteqn}) explicitly affords

\begin{eqnarray}
\psi(x,t) &=& \frac{1}{\sqrt{2\pi}} \left(\frac{\sigma^2}{\pi}\right)^{1/4} \int  e^{(k-k_0)^2\sigma^2 / 2} e^{-ika}  e^{-ik^2 t} C(k) e^{ikx} \, dk \\
&=& \int \left(\frac{1}{\sqrt{2\pi}}e^{ikx} \right) \phi'(k,t) \, dk,
\end{eqnarray}

\noindent where in the second line $\psi(x,t)$ was rewritten in terms of the Fourier transform of a function $\phi'(k,t) = C(k) \exp [-ik^2 t] \phi(k,0)$ comprising a Gaussian envelope with amplitude $C(k)$ inherited from the plane wave.  Using this representation, the norm of the packet is simply

\begin{eqnarray}
\mathcal{N} = \vert\vert \phi'(k,t) \vert\vert^2 &=& \int [\phi'(k,t)]^* \phi'(k,t) \, dk \\
&=& \left(\frac{\sigma^2}{\pi}\right)^{1/2} \int e^{(k-k_0)^2\sigma^2 } \vert C(k) \vert^2 \, dk.
\end{eqnarray}

\noindent Assuming unit incident norm, the norm of the transmitted or reflected packet equates to the transmission or reflection amplitude, respectively.  To compute this explicitly, let a packet be incident on the right side of the $\mathcal{PT}$ symmetric region ($B_R = 1$ and $A_L = 0$) so that the amplitude of the reflected wave is $A_R = M_{12}(k) / M_{22}(k)$.  Then the reflection coefficient for the wave
on the right side is given by

\begin{equation}
R_R =   \left(\frac{\sigma^2}{\pi}\right)^{1/2}\int e^{(k-k_0)^2\sigma^2} \left| \frac{M_{12}(k)}{M_{22}(k)} \right|^2 \, dk,
\end{equation}

\noindent and, with $B_R = 1 / M_{22}(k)$, we have the transmission coefficient

\begin{equation} \label{wptranscoeff}
T_R =  \left(\frac{\sigma^2}{\pi}\right)^{1/2}\int e^{(k-k_0)^2\sigma^2 } \left| \frac{1}{M_{22}(-k)} \right|^2 \, dk,
\end{equation}

\noindent with the  sign change due to the opposite motion of the plane wave, though this is strictly formal since $\vert M_{22}(k)\vert$ is an even function of $k$ for the given potential.  Note that these integrals are well defined with a removable singularity at $k = 0$.  The accuracy of this framework requires that the barrier width and phase factor are suitably chosen so that the packet does not spread appreciably on the traversal timescale.  One caveat of this analysis is that the reflected packet must maintain a Gaussian profile; a condition which is only satisfied for a certain range of parameters due to the saturation of anisotropic transmission resonances.

\subsection{Potential Structure}

\par
\begin{figure} 
\centering
 \includegraphics[scale=1.0]{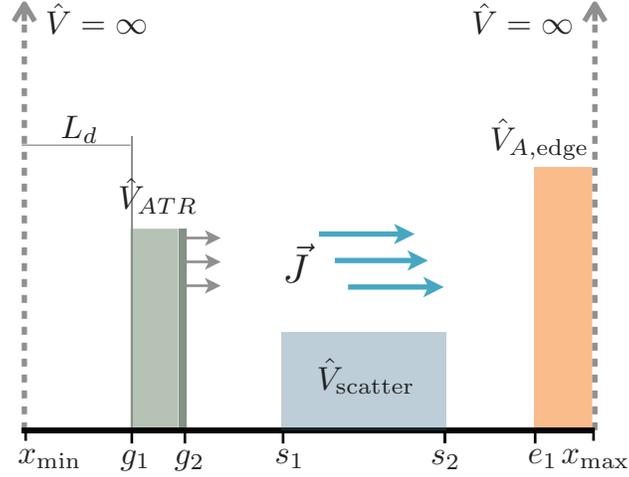} 
\caption{Cross--sectional geometry for wavepacket propagation with boundary wavepacket generation.  The interaction region with scattering potential $\hat{V}_\text{scatter}$ is  situated between an absorbing potential $\hat{V}_{A,\text{edge}}$ and a  generating potential $\hat{V}_{ATR}$.  The edge absorbing potential $\hat{V}_{A,\text{edge}}$ completely attenuates any wavepacket that enters this region, while the $\mathcal{PT}$--symmetric ATR potential $\hat{V}_{ATR}$ has a generating surface oriented toward the scattering region.  Any wavepacket that crosses the ATR edge causes a new counter--propagating packet to be reflected, while itself passing through the potential and reflecting off the wall of the infinite square well.}
\label{edgegen_geom}
\end{figure}

 Having developed a toolkit containing both absorbing and generating potentials, these components may be assembled to afford an effective simulation method for open systems.  The most intuitive construction entails placing an edge absorbing potential $\hat{V}_{A,\text{edge}}$ at the boundary of the simulation cell, which is assumed to lie within an infinite square well, and an ATR generating potential $\hat{V}_{ATR}$ near the other boundary [Fig. \ref{edgegen_geom}].  The generating face of the absorbing potential is oriented toward the scattering region, so that any particle incident on this region will transmit and be compensated by an additional reflected particle.  The transmitted particle will ultimately reflect off the square well boundary at $x_\text{min}$ and reenter the system. Within the center of the cell these wavepackets encounter a scattering region in which the particles interact with static potentials or through many--body interactions.  After traversing this region, the particle reaches $\hat{V}_{A,\text{edge}}$, where it is completely attenuated.  This establishes a net current from the generating region to the absorbing region.  Note that neither the generating nor attenuating potentials overlap with the scattering region, ensuring that the transport processes are unperturbed.

\par In this scenario, the absorbing potential $\hat{V}_{A,\text{edge}}$ is chosen so that any packet entering this region is completely attenuated before reaching the edge of the square well.  If the potential is sufficiently strong, reflections and transmission will be minimized, thereby eliminating a source of artifacts.  The definition of the ATR potential is slightly more complicated.  Due to the saturable reflections inherent in ATR potentials,  the generated packet will only be Gaussian (and not an extended Gaussian), if the width of the incident wavepacket is greater than the region $L_\text{ATR} = \vert g_1 - g_2 \vert$ over which $\hat{V}_{ATR}$ is defined.  Furthermore, the width of this region and the distance $L_d = \vert x_\text{min} - g_1 \vert$ determine the interpacket spacing or the delay time $2 (L_d + L_\text{ATR}) / v_g$ between packet arrivals.  Note that, by construction, this method requires the density from within the scattering region to impinge on the generating potential in order to afford a positive flux of  norm.  Thus, the bias across the simulation must be suitably small so that backscattered packets  continue to reach  $\hat{V}_{ATR}$.  For steady--state current (no time--dependent potentials or charge accumulation) we require that the total norm within the cell remain constant at all times, and hence $\langle\partial \mathcal{N}_G / \partial t\rangle = -\langle\partial \mathcal{N}_A / \partial t\rangle$, where $\mathcal{N}_G$ is the generated norm and $\mathcal{N}_A$ is the attenuated norm for the system.

\par
\begin{figure} 
\centering
 \includegraphics[scale=1.0]{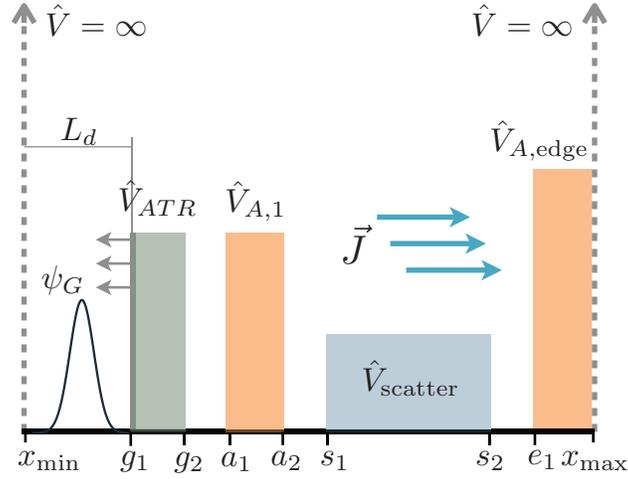} 
\caption{Cross--sectional geometry for wavepacket propagation from an ATR pulse train generator.  The interaction region with scattering potential $\hat{V}_\text{scatter}$ is  situated between an absorbing potential $\hat{V}_{A,\text{edge}}$ and the pulse generator comprising an ATR potential $\hat{V}_{ATR}$, a Gaussian absorbing potential $\hat{V}_{A,1}$ and the seed wavepacket $\psi_G (x,t)$.  The reflecting surface of the ATR potential faces $\psi_G(x,t)$, ensuring a packet will remain in the generator while affording a pulse stream toward the interaction region. }
\label{bounce_geom}
\end{figure}

 A second scenario may be envisioned, in which a generating potential consistently adds a stream of packets with fixed delay spacing to the system.  In such a configuration, the outgoing particles are once again attenuated by a potential at the cell boundary $\hat{V}_{A,\text{edge}}$, however, a second absorbing potential $\hat{V}_{A,1}$ and generating potential $\hat{V}_{ATR}$ are utilized to create a pulse generator [Fig. \ref{bounce_geom}].  Specifically, a wavepacket $\psi_G(x,t)$ of norm $\braket{ \psi_G(x,t) \vert \psi_G(x,t)} = \mathcal{N}_g$ with $\mathcal{N}_g > 1$ is placed between the ATR and the cell boundary, with the generating face of the ATR facing toward the cell edge. During simulation, the packet $\psi_G(x,t)$ impinges on the reflecting surface of the ATR causing an identical packet to be reflected into the delay region accompanied by transmission of the incident packet toward the scattering region.  The transmitted packet then passes through $\hat{V}_{A,1}$, which attenuates the particle to unit norm before it interacts with the scatterers.  In the same manner, $\hat{V}_{A,1}$ attenuates any packets passing from this interaction region toward the pulse generator, isolating it from the simulation.  The new packet generated in the delay region reflects off the cell boundary at $x_\text{min}$ and propagates back toward the generating surface at $g_1$ to begin this process anew.  The use of pulse trains generated in this manner is particularly appealing for situations where non--equilibrium charge accumulation, $\langle\partial \mathcal{N}_G / \partial t\rangle \neq -\langle\partial \mathcal{N}_A / \partial t\rangle$ is desirable such as in capacitive charging.

\par In the packet generator configuration, the pulse generation delay time is given by $2 L_d / v_g$, and hence is a tunable parameter.  The norm $\mathcal{N}_g$ of the generating packet must be adjusted for the given absorbing potential $\hat{V}_{A,1}$, as the attenuation rate is a function of both the potential itself and the norm of the incident wavefunction (Eq. \ref{attenrate}).  Note that the attenuation rate is larger for a packet with a larger norm, and thus the rate of absorption for a reflected packet incident from the scattering region will be less than that for a probe packet incident from the packet generator.  The applicable timescale for this method is likewise limited by the scheme utilized to maintain the wavepacket(s) in the generating region, as they will ultimately broaden under time evolution in the absence of measurement.

In this scheme, a complication regarding transferability to different biases results from the use of generating potentials.  At a finite bias voltage $V$ the energy $E_0$ of a given particle will undergo a shift to $E = E_0 + eV$.  This corresponds to a new wavevector $k = \sqrt{2m(E_0 + eV)}$, and hence a new group velocity for the packet $v_g = \sqrt{2m(E_0 + eV)} / m$.  Thus, the absorbing potential parameters  need to be reoptimized at each finite bias, or the bias range chosen to be sufficiently narrow, to ensure the addition of unit norm packets with minimal reflection.  This is less of a concern for the boundary absorbing potential, as the strength and width may be initially chosen so as to attenuate any incident packets for a range of energies $E_0 \pm eV$.  Nonetheless, the widths of absorbing and generating regions must be altered for both propagation schemes since the shift in group velocity affects the extent of norm generation or loss. Specifically, the net norm removed from the system is
\begin{eqnarray}
\mathcal{N}_{A} &=& \int_{t_i}^{t_f} dt  \,\frac{\partial{N}_{A}}{\partial t} \\
&=& \int_{t_i}^{t_f} dt \, \braket{ \psi(\vec{x},t) \vert \text{Im}(\hat{V}_{PT})\vert \psi(\vec{x},t)} \\
&=& \int_{t_i}^{t_f} dt \int_\mathcal{V} d^N x \, \psi^*(\vec{x},t)\, [\text{Im}(V_{PT}(\vec{x}))] \, \psi(\vec{x},t) \, 
\end{eqnarray}

\noindent such that $\Delta t = t_f - t_i = t_S$ = $L / v_g$ is the duration for which the attenuating potential acts, $\mathcal{V}$ is the volume of the absorbing region, and $N$ is the dimensionality of the system. The delay region must likewise be modified to ensure a proper inter--packet delay time.   Finally, an ultimate time scale must be assigned to the stability of these simulations due to aberrant accumulation or loss of norm.  This deviation may arise from either imperfect transmission, reflection, and generation, or from the inevitable spread of the generating wavepacket.

\section{Numerical Results}

\subsection{Propagation Parameters}

\par Numerical simulations are performed through real--time propagation of an initial wavepacket.  The propagation method, detailed in the Appendix, employs a forward finite difference algorithm to propagate real and imaginary components of the wavefunction. The initial wavepacket is described by the product of a normalized, unit mass Gaussian centered at $x_0$ and a monochromatic plane wave as

\begin{equation}
\psi(x,0) = \frac{1}{(\pi \sigma^2)^{1/4}} e^{-(x-x_0)^2 / 2\sigma^2} e^{ik_0 (x-x_0)}
\end{equation} 

\noindent where $k_0 = \sqrt{2E}$ is the initial wavevector for a particle of energy $E$ and $2\sigma \sqrt{2 \log 2}$ is the full--width at half--maximum (FWHM) spatial extent of the packet.  The packet propagates in the direction of $k_0$ with frequency $\omega = k^2 /2$ and group velocity $v_g = \partial \omega / \partial k = k$. The wavepacket is discretized on a spatial lattice comprising $N = 1 \times 10^4$ elements and integrated with finite temporal and spatial steps, $\Delta t = 1.0 \times 10^{-9}$ and $\Delta x = 1.0 \times 10^{-4}$ respectively.  This ensures that the lattice spacing is  smaller than the phase oscillation length of the packet for a typical choice of parameters ($\sigma^2 = 0.001$ and $k_0 = 500$).  Arbitrary potentials are defined within the confines of the lattice, with infinite square--well boundary conditions ensuring that the wavefunction vanishes at the edges of the cell.

\subsection{ATR Potential Numerics}

\par
\begin{figure} 
\centering
 \includegraphics[scale=1.0]{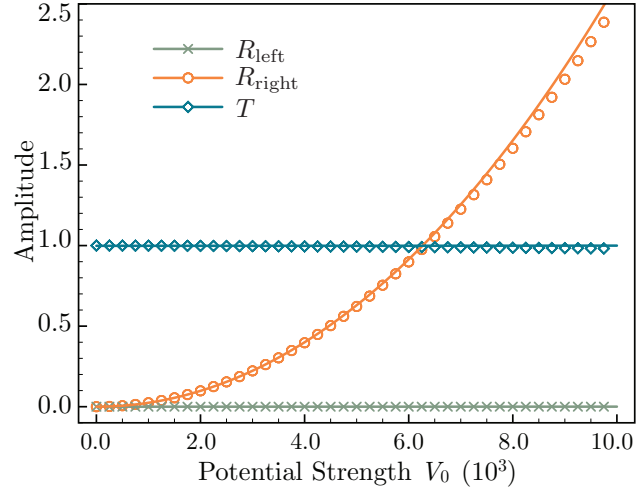} 
\caption{Analytical (solid lines) and simulated (points) transmission $T$, standard reflection $R_\text{left}$ and enhanced reflection $R_\text{right}$ coefficients for passage of a wavepacket ($k_0 = 500$, $\sigma^2 = 0.001$) through an ATR potential region ($L_\text{ATR} = 20 \Lambda$) as a function of the potential strength $V_0$.}
\label{coefficients}
\end{figure}

 Scattering from an ATR potential was simulated via real--time propagation of an initial Gaussian wavepacket ($k_0 = -500$, $\sigma^2 = 0.001$, $x_0 = 0.80$) incident on an ATR region of width $L = 20 \Lambda = \pi / 25$ centered about $x = 0.50$.  The scaling behavior of the reflection and transmission coefficients exhibits good agreement with analytical calculations, with a few notable deviations [Fig. \ref{coefficients}].  In particular, the right (enhanced) reflection coefficient $R_\text{right}$ and the transmission coefficient $T$ are found to be nontrivially smaller than the analytical result when the ATR potential strength is greater than $V_0 \sim 6.0 \times 10^{-3}$.  This corresponds to a regime for which $R_\text{right} > 1.0$, and hence where the wavefunction norm is doubled.  The discrepancy may arise from the approximation of eigenfunctions within the ATR region as modified Bessel functions of the first kind, and thus the neglect of Jordan associated functions.  Additional deviations are due to the spread of the wavepacket during propagation, as the FWHM no longer corresponds to that defined by $\sigma^2 = 0.001$ in the initial distribution.  Nonetheless, calculations in which  wavepacket propagation was initiated as close as possible to the ATR region demonstrate that violations of the quasistatic approximation arising from wavefunction spread do not account for these large discrepancies in the data.

\par
\begin{figure} 
\centering
\includegraphics[scale=1.0]{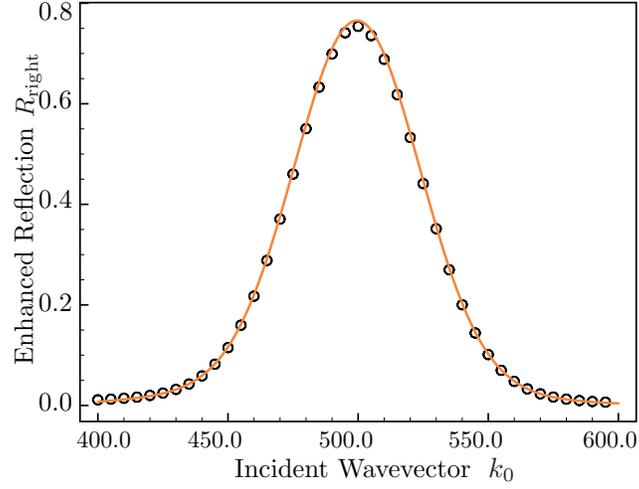} 
\caption{Analytical (solid lines) and simulated (points) scaling of the enhanced reflection coefficient $R_\text{right}$ as a function of incident wavevector $k_0$ for a Gaussian wavepacket ($\sigma^2 = 0.01$) impinging on an ATR potential ($V_0 = 5500$, $L_\text{ATR} = 20 \Lambda$). }
\label{enhanced_vs_k}
\end{figure}

 There is a strong dependence of the enhanced reflection coefficient on the incident wavevector when scattering from a grating with fixed ATR mode wavevector $\Lambda = \pi / k_\text{grating} = \pi / 500$ [Fig. \ref{enhanced_vs_k}].  Nonetheless, the reflection coefficient $R_\text{right}$  is reduced by a factor of $0.90$ for wavevectors $k_0 = 500 \pm 10$, corresponding to incident packet energies ranging between $E = E_0 \pm 5000$.  Thus, if used as a generating potential in transport calculations, this ATR configuration would ensure greater than 90.0\% generation for bias values of $eV = \pm 5000$, or $\sim 4.0 \%$ of the incident packet energy.  Such a dispersion is more than suitable for most transport applications, in which the bias need not exceed a few electron volts.

\par
\begin{figure} 
\centering
 \includegraphics[scale=1.0]{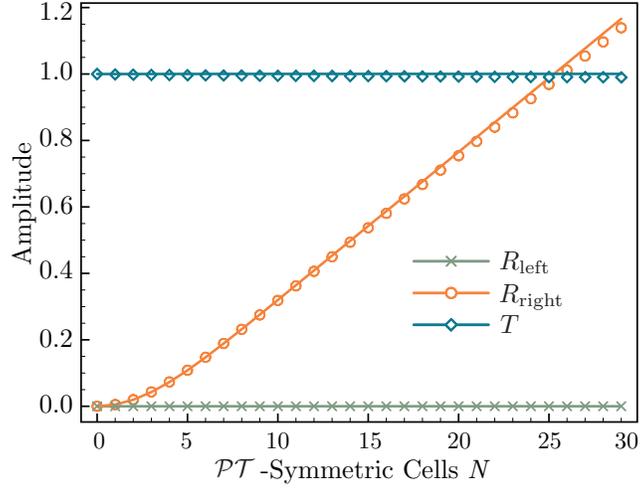} 
\caption{Analytical (solid lines) and simulated (points) transmission $T$, standard reflection $R_\text{left}$ and enhanced reflection $R_\text{right}$ coefficients for passage of a wavepacket ($k_0 = 500$, $\sigma^2 = 0.001$) through an ATR potential region of variable length $L = N\Lambda$ where $\Lambda = \pi / 500$.}
\label{length_dependence}
\end{figure}

 The enhanced reflection coefficient ($R_\text{right}$) is found to exhibit an initial quadratic dependence on the number of $\mathcal{PT}$--symmetric ATR unit cells, followed by a linear increase at cell numbers $N \geq 5$ [Fig. \ref{length_dependence}].  The transmission coefficient drops below unity for large ATR crystals, however, the overall magnitude of this effect is rather small ($T \sim 0.989$ at $N = 30$).  For simulation purposes, it is desirable to keep the length of the ATR region smaller than the width of the Gaussian to prevent extrusion of the generated packet.  For $\sigma^2 = 0.001$, which represents a rather broad packet, this requires $N \leq 20$.  

\par
\begin{figure} 
\centering
 \includegraphics[scale=1.0]{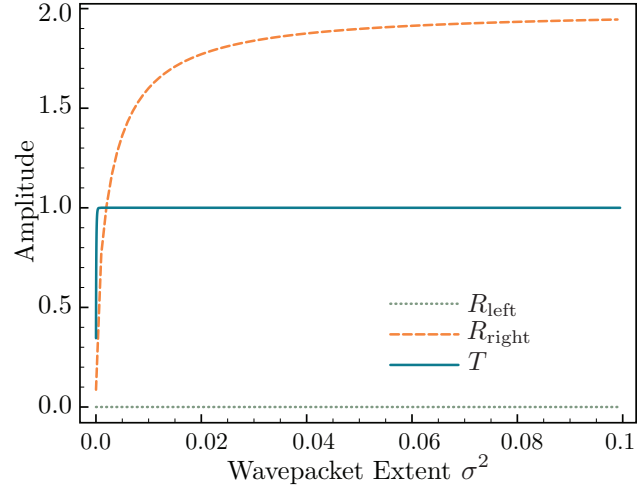} 
\caption{Calculated  transmission $T$, standard reflection $R_\text{left}$ and enhanced reflection $R_\text{right}$ coefficients for passage of a Gaussian wavepacket ($k_0 = 500$) through an ATR potential region of length $L = 20\Lambda$ and $V_0 = 5500$ as a function of the wavepacket extent $\sigma^2$.}
\label{sgsq_dependence}
\end{figure}

 The dependence of transmission properties on $\sigma^2$ is important for the stability of a packet generator, yet this is difficult to quantify numerically due to spread of the packet during real--time propagation.  Using the analytical results as a guide, a strong dependence exists between the enhanced reflection coefficient $R_\text{right}$ and the incident packet width [Fig. \ref{sgsq_dependence}].  As the packet broadens spatially under time evolution, $\sigma^2 \longrightarrow \infty$, the momentum distribution narrows and $R_\text{right}$ asymptotically approaches the value expected for an incident plane wave.  Thus, if a broad initial packet is chosen, there will be little change in the enhanced reflection coefficient as a function of time, leading to a longer timescale for stable packet generation.  Conversely, if a narrow initial packet is chosen, the enhanced reflection coefficient will vary substantially between subsequent generation events as $\sigma^2$ grows, lending inconsistency to the simulation.

\subsection{Transport Through A Scattering Region}

\par Calculations employing the ATR generator were performed using broad wavepackets $\sigma^2 = 0.01$ with large norm $\mathcal{N} = 8$ and an incident wavevector $k = 500$ corresponding to an energy of $E = 1.25 \times 10^5$.  The generating wavepacket was situated between the edge of the simulation cell and  an ATR generating potential of width $L = 5 \Lambda = \pi / 100$.  The potential was numerically optimized to yield $V_0 = 16049.5$, which affords unit generation and transmission of the incident packet. A Gaussian absorbing potential was situated between the ATR and the scattering region, and defined over a distance $L_\text{Gau} = 0.10$ with $\alpha^2 = 1.0 \times 10^{-4}$.  The magnitude of the absorber was numerically optimized to yield $V_\text{Gau} = 520.1$, which attenuates an incident $\mathcal{N} = 8$ wavepacket to unit norm.  The use of a large incident packet permits a large Gaussian filter, which reduces the penetration of reflected wavepackets into the ATR region.  Wavepackets were attenuated at the outgoing boundary of the scattering region using a singular absorbing potential with $c = 2.0$,  $k_\text{min} = 250$, and a width $L_\text{att} = 0.250$.  The scattering region was occupied by a rectangular potential barrier $L_\text{barrier} = 0.095$ units in extent and evaluated at a variety of potential strengths $V_\text{barrier}$ to determine conductance characteristics.  All components were enclosed in an infinite square well measuring 2.0 units in spatial extent [Fig. \ref{bounce_geom}].  The spatial integration step was taken to be $\Delta x = 2.0 \times 10^{-4}$ in this case.

\par
\begin{figure} 
\centering
 \includegraphics[scale=1.0]{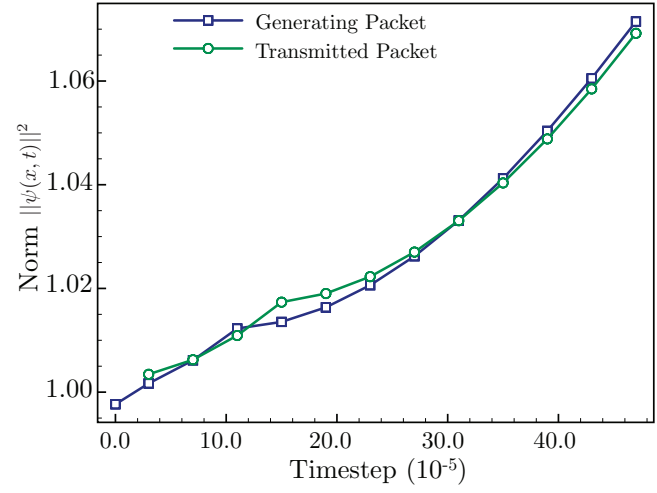} 
\caption{Evolution of  wavefunction norm for generated (reflected) and transmitted packets arising from the ATR packet generator during successive enhanced reflection events.}
\label{stability}
\end{figure}

\begin{figure} 
\centering
 \includegraphics[scale=1.0]{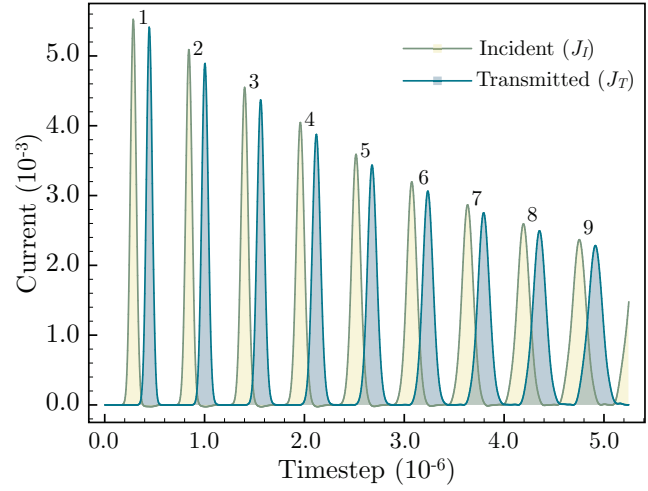} 
\caption{Incident $J_I$ and transmitted $J_T$ current arising from an ATR potential and incident on a rectangular barrier with $V_\text{barrier} = 2.5 \times 10^4$ and $L_\text{barrier} = 0.1$ units.  The initial generating packet is a Gaussian with wavevector $k_0 = 500$ and $\sigma^2  = 0.01$ characterizing the packet width.}
\label{successive}
\end{figure}

 Calculations performed with $V_\text{barrier} = 0$ reveal that the norms of the generating and transmitted packets are well maintained, with a deviation of $\sim$ 10\% observed after generation of twelve packets, corresponding to over $5.0 \times 10^6$ integration timesteps [Fig. \ref{stability}].  This is comparable to the divergence expected for the simple first--order integration scheme employed herein.  Accordingly, the only factor that varies substantially between subsequent iterations aside from this systematic error is the spread of the wavepacket.  To ascertain the current flow through a scattering region, the probability current was averaged over two small spatial windows, measuring 0.04 units in width, placed on either side of the rectangular barrier and a bias potential $V_\text{bias}$ was added to the incident packet $E = E_{k_0} + V_\text{bias}$.  This configuration conceptually resembles a conventional four--probe conductivity measurement.  The conductance $\mathcal{G}$ of the scattering region at a given bias $eV$ may be obtained as a function of the transmission coefficient $T$:

\begin{eqnarray}
\mathcal{G}(eV) = \frac{e^2}{\pi \hbar} T(eV) = \frac{e^2}{\pi \hbar} \frac{\vert J_T(eV) \vert}{\vert J_I (eV) \vert}
\end{eqnarray}

\noindent where $J_I$ is the incident wavepacket current and $J_T$ is the transmitted wavepacket current \cite{Imry1999}.  The spread of the wavepacket in the ATR region has a demonstrable effect on successive transmitted packets as measured at zero applied bias, which manifests through a decrease in the peak current density [Fig. \ref{successive}].  Nonetheless, the conductance values remain remarkably stable even as the barrier strength is increased, with the first four transmission events affording nearly identical conductance determinations (Table \ref{conduct}).  When including the full set of nine transmission events the calculated conductance varies by only 6.4\% of the value calculated from the first event.  As a point of reference, the conductance was analytically determined using the transmission coefficient for a plane wave through a square barrier

\begin{equation}
T = \left( 1 + \frac{V_\text{barrier}^2 \sin^2 (k L_\text{barrier})}{4E(E-V)}\right)^{-1}
\end{equation}

\noindent in conjunction with Eq. (\ref{wptranscoeff}).  In this context, $V_\text{barrier}$ is the barrier height and $L_\text{barrier}$ the barrier width, and $k = \sqrt{2E}$ is the incident wavevector. The analytically--determined values agree closely with those obtained numerically for low barriers, with a slight departure from analytical results in the high--barrier case.  This discrepancy likely arises due to deviation of the generated packet shape from a proper Gaussian, made more apparent due to reflection from a stronger potential.  In either case, the magnitude of this deviation never exceeds 10\% of the analytical value affording an accuracy beyond other numerical schemes for conductance determination (Table \ref{conduct}).

\par
\begin{figure} 
\centering
 \includegraphics[scale=1.0]{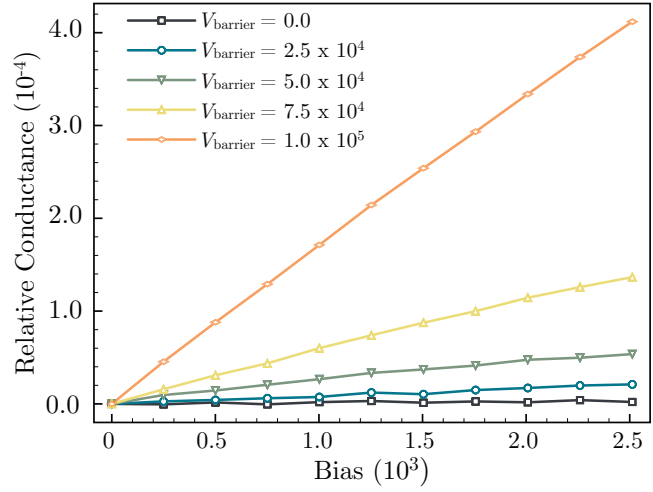} 
\caption{Relative conductance $(\mathcal{G}(V) - \mathcal{G}(0)) / (\mathcal{G}_0(V) - \mathcal{G}_0(0))$  calculated as a function of bias $E = E_{k_0} + V_\text{bias}$ for several rectangular barrier strengths $V_\text{barrier}$.  The initial generating packet is a Gaussian with $k_0 = 500$ as the initial wavevector and $\sigma^2  = 0.01$ characterizing the packet width, while the scattering potential has an extent of $L_\text{barrier} = 0.1$ units.}
\label{ivc}
\end{figure}

 The ATR packet generating scheme employed herein is essentially a response formalism, in which the reaction of a system to a probe packet is measured.  Accordingly, there exists a nonzero current at zero applied bias, which comprises the reference state for such determinations.  Physically, the zero--bias state in a material is associated with zero net current, and hence an isotropic movement of charge carriers in the system.  The formalism herein corresponds to the short time limit, in which a single carrier has passed in a given direction but before an additional compensatory carrier may pass in the opposite direction.  To demonstrate the scaling of transport with applied bias, it is more instructive to consider the relative conductance versus bias than the raw transmitted current.  The relative conductance $\mathcal{G}_\text{Rel} (V_\text{bias})$ is defined as

\begin{eqnarray}
\mathcal{G}_\text{Rel}(V_\text{bias}) &=& \frac{\mathcal{G}(V_\text{bias}) - \mathcal{G}(0)}{\mathcal{G}_0 (V_\text{bias}) -\mathcal{G}_0 (0)} \\
&=& \left. \left(\frac{J_T(V_\text{bias})}{J_I(V_\text{bias})} - \frac{J_T(0)}{J_I(0)}\right)  \middle/ \left(\frac{J_{0,T}(V_\text{bias})}{J_{0,I}(V_\text{bias})} -\frac{J_{0,T}(0)}{J_{0,I}(0)}\right) \right.
\end{eqnarray}

\noindent where a subscript of zero indicates the current or conductance calculated in the absence of a barrier.  The normalization of the transmitted current by the incident current is required for comparative purposes between calculations with different barriers, as the presence of the barrier itself introduces a boundary condition which may alter the incident flux.    Furthermore, as all determinations are taken with respect to a probe packet, the conductance must be measured relative to that observed in the absence of a barrier to provide a reference point for free propagation and accommodate variation in peak--to--peak current due to packet spread.   The result of this analysis is in some sense analogous to the I--V curves typically presented in the context of experimental transport measurements.  The scaling of the relative conductance $\mathcal{G}_\text{Rel}(V)$ exhibits the expected correlation with increasing bias and increasing barrier strength [Fig. \ref{ivc}].  Notably, the increase in barrier strength affords a greater slope for $d\mathcal{G}_\text{Rel}(V) / dV_\text{bias}$, consistent with the expected scaling for the transmission coefficient through an increasingly strong rectangular barrier.  It is notable that the formalism herein affords the conductance at both zero and finite bias with no additional computational cost.

\section{Computational Limitations of Complex Potentials in Many--Body Systems}

\par While evolution under the action of $V_{PT}$ mimics a multiparticle state, this does not embody all the requisite properties for a true many--body configuration. To see this, assume a simple system with a wavefunction given by the product ansatz $\ket{\Psi(t)} = \ket{\psi_0 (t)} \otimes \ket{\psi_1(t)}$, where $\ket{\Psi(t)} \in \mathcal{H}^{(2)} = \mathcal{H} \otimes \mathcal{H}$ is a two--particle Hilbert space.  For now we ignore the effects of symmeterization, as this elementary form is sufficient for illustrative purposes.  The full Hamiltonian for this system is

\begin{equation}
\hat{H} = \hat{H}_0 \otimes \hat{H}_0 + \hat{V}_{A/G} \otimes \mathbb{I} + \hat{V}_{2p}
\end{equation} 

\noindent where $\hat{H}_0$ is the Hamiltonian for an isolated particle, $\hat{V}_{A/G}$ is the complex potential term acting only on $\ket{\psi_0 (t)}$, and $\hat{V}_{2p}$ is a two--particle interaction defined by

\begin{eqnarray} \label{twobody}
\hat{V}_{2p} &=& \sum_{ij}  (\ket{\psi_i} \otimes \ket{\psi_j}) V_{2p}^{ij} (\bra{\psi_i} \otimes \bra{\psi_j}) \\
&=& \sum_{ij}  (\ket{\psi_i} \otimes \ket{\psi_j}) U_{2p}^{ij} (\delta_{ij} - 1) (\bra{\psi_i} \otimes \bra{\psi_j}) 
\end{eqnarray}

\noindent which approximates the Hartree--like term in an electronic structure method. Assume once again that $\hat{V}_\text{A/G}$ may be turned on or off arbitrarily, or asymptotically localized to a region of space, so that the interaction will apply to  $\ket{\psi_0 (t)}$ only when it traverses this region.  The latter scenario is representative of the complex absorbing and generating potentials utilized herein. To further simplify discussion, take $\hat{V}_\text{A/G}$ to be entirely imaginary, as the real part of this potential may be absorbed into $\hat{H}_0$ as a single--particle potential term. The time evolution operator decomposes as a tensor product in this formalism 

\begin{equation}
\hat{U} = (\hat{U}_0 \otimes \hat{U}_0) (\hat{U}_{A/G} \otimes \mathbb{I})
\end{equation}

\noindent where $\hat{U}_0(t_2,t_1) = \exp[-i(\hat{H}_0 + \hat{V}_{2p}) (t_2 - t_1) /\hbar]$ is the evolution in the absence of the complex potential and $\hat{U}_{A/G} =  \exp[-i\hat{V}_{\text{A/G}} (t_2 - t_1) / \hbar]$ is the nonunitary evolution afforded by the Hermicity breaking term.  

\par Assume that $\ket{\Psi(t)}$ evolves in the absence of $\hat{V}_\text{A/G}$ up to a time $t_1$ after which  $\ket{\psi_0(t_1)}$ enters the interaction region. Furthermore, let the interaction with $\hat{V}_\text{A/G} \ket{\psi_0(t_1)} = i\Gamma \ket{\psi_0(t_1)}$ end at $t_2$ sometime later.  During this propagation, the wavefunction is carried to the final state $\ket{\Psi(t_2)} = \ket{\psi'_0 (t_2)} \otimes \ket{\psi_1(t_2)}$, where $\ket{\psi'_0 (t_2)} = \exp[\Gamma (t_2 - t_1) / \hbar] \ket{\psi_0 (t_2)}$, so that $\ket{\psi_0 (t_2)}$ corresponds to time evolution under the Hermitian part of the Hamiltonian.  Defining $\alpha = \exp[2\Gamma (t_2 - t_1) / \hbar]$, it is clear that $0 \leq \alpha \leq 1$ for an attenuating potential $\hat{V}_A$ and $1 \leq \alpha < \infty$ for a generating potential $\hat{V}_G$.  Focusing on the latter case, it is desirable to choose $\hat{V}_G$ such that $\alpha \in \mathbb{N}$, thereby ensuring that norm generation occurs in units of a single particle.   If the norm of $\ket{\psi'_0 (t_2)}$ is enhanced to correspond to a two--particle state ($\alpha = 2$), then the interaction with $\ket{\psi_1 (t_2)}$ is scaled accordingly as

\begin{equation}
\braket{\Psi(t_2) \vert \hat{V}_{2p} \vert \Psi(t_2)} = 2 U_{2p}^{01}
\end{equation}

\noindent which corresponds to the doubling of the potential term due to the interaction of a single particle in $\ket{\psi_1 (t_2)}$ with the two ``particles'' in  $\ket{\psi'_0 (t_2)}$.  This result is not physically meaningful, as the system is now analogous to a three--particle problem in which two of the particles interact with the third particle, but not with each other.  The origin of this fault  arises from the nature of the generating potential itself, which superposes  additional norm onto an existing wavefunction instead adding an additional state vector as required for a true multiparticle configuration.  Since the net effect of the complex potential is only to elongate a state vector and not to create a new state, the presence of a nonzero coupling term for the new particle and its parent can only be achieved by artificially introducing a self--interaction term in the Hamiltonian.  Nonetheless, for any conventional two--body potential, however, the vanishing diagonal terms in Eq. (\ref{twobody}) will prevent these states from acting as a true multiparticle configuration.  These observations collectively impose strong limitations on the computational scope of any calculation that employs complex generating potentials.  Specifically, the use of generating potentials excludes any wavefunction based method, or any method that includes Hartree--Fock exchange, from consideration in this context.

%%
%% Is sum and dilation the correct verbiage here?  and is this actually c\psi(kx - d)?
%%

\par These limitations may be circumvented through the use of  theories that are formulated in terms of the norm of constituent states, such as DFT.  The DFT Hamiltonian is defined solely in terms of the single--particle density $\rho(x)$ such that 

\begin{equation}
\rho(\vec{x}) = N \int d^3 \vec{x}_2 \dots d^3 \vec{x}_N \vert \Psi_0(\vec{x}, \vec{x}_2, \dots, \vec{x}_N) \vert^2
\end{equation}

\noindent where $\Psi_0(\vec{x}, \vec{x}_2, \dots, \vec{x}_N)$ is the $N$--particle ground--state wavefunction characterizing the system.  In this scheme, terms that are pathological for generating potential--modified wave functions, such as the Hartree interaction

\begin{equation}
V_\text{Hartree}[\rho(x)] = \frac{e^2}{2} \int\int d^3 \vec{x} \, d^3 \vec{x}'\, \frac{\rho(\vec{x})\rho(\vec{x}')}{\vert\vert \vec{x} - \vec{x}' \vert\vert} 
\end{equation}

\noindent cease to be problematic as there is no explicit dependence on single--particle state vectors.  The role of an absorbing or generating potential is then to modulate $\rho(x)$ in a manner that adds density to or subtracts density from the system.  Note that these considerations apply only to pure DFT. Hybrid methods, which incorporate a degree of exact exchange from Hartree--Fock theory, will suffer from the same failures as full wavefunction methods.  

\par The results for the propagation of a single wavepacket considered in this manuscript are directly applicable to DFT by construction.  In the single--particle limit, the particle density from DFT reduces to $\rho(x) = \psi^*(x) \psi(x)$, and thus the rescaling induced by the absorbing or generating potential transforms the density in a manner identical to the wavefunction norm discussed herein.  Furthermore, the choice of Gaussian wavepackets underscores the correspondence with DFT, in which Gaussian functions are a popular functional form in localized and hybrid localized/delocalized basis set schemes.  Thus, the single--packet simulations are analogous to a valence electron traversing the system boundaries in the limit of vanishing coupling to the other electrons and ions.

\section{Conclusions}

\par The computational methods developed herein outline a path through which $\mathcal{PT}$--symmetric potentials may be employed to afford open boundary conditions in the context of RT--TTDFT transport calculations.  Existing methods have utilized absorbing boundary conditions to attenuate wavefunction norm at simulation boundaries, however, this does not permit the complementary positive probability density flux  required for a physically realistic system.  A judicious assembly of ATR regions permits construction of a wavepacket pulse generator that can inject a train of probe wavepackets into the scattering region.  By measuring the ratio of outgoing to incoming current, the transmission coefficient and hence conductance are calculated as those of a single conducting channel \cite{Imry1999}.  As an ancillary benefit, the zero--bias and finite--bias conductance may be readily determined in the presence of time dependent processes including, but not limited to, the oscillatory electric fields associated with  photoexcitation.  This  transport formalism is demonstrated to exhibit excellent agreement with analytical results, paralleling  the recent success using similar $\mathcal{PT}$--symmetric methods to describe open quantum dots \cite{Berggren2010, Wahlstrand2014}, dipolar Bose--Einstein condensates in open double--well potentials \cite{Fortanier2014}, and the topologically trivial and nontrivial phases of the Su--Schrieffer--Heeger model with open chain boundaries \cite{Zhu2014}.  A particular property of $\mathcal{PT}$--symmetric Hamiltonians prevents these methods from generating a true many--body wavefunction and hence this formalism is not applicable to Hartree--Fock  or explicit multireference methods.  Nonetheless, these limitations do not apply to modulation of the probability density, so that $\mathcal{PT}$--symmetric potential terms may be employed without restriction in any DFT--based formalism.

\par This method is numerically robust, exhibiting stable transmission characteristics for up to nine transfer events in a simple model system.  This exceeds the timescale accessible through prior real--time propagation calculations by several orders of magnitude, in which only a fraction of a carrier may be transferred before the simulation becomes unstable due to carrier depletion \cite{Varga2011}.  Furthermore, the temporal upper limit for RT--TDDFT calculations in actual materials is limited by the highest phonon frequency of the material.  On this timescale, the lattice undergoes spatial translation, electron--phonon coupling terms become nontrivial, and the adiabatic approximation ceases to hold.  This corresponds to only a few carrier transfer events.  Thus, the framework herein affords boundary conditions for RT--TDDFT throughout its range of physical applicability. 

\par Nonequilibrium Green's function methods currently comprise the mainstay for explicit quantum transport calculations, although time--dependent phenomena are inaccessible in this context due to their static nature.  Conductances calculated in this scheme likewise deviate from experimentally determined values by one to two orders of magnitude, limiting this method to use as a qualitative tool that indicates physical mechanism through scaling behavior.  RT--TDDFT ameliorates the restrictions imposed by the quasi--static approximation, while affording conductance values within 10\% of analytical results for a model system.  Thus the conjunction of RT--TDDFT with ATR potentials is a firm step toward the development of broadly applicable and quantitatively accurate electronic structure methods for quantum transport in real materials.

\section{Acknowledgements}

\par This research was supported by the start--up grant and University Facilitating Fund of George Washington University.  Computational resources utilized in this research were provided by the Argonne Leadership Computing Facility (ALCF) at Argonne National Laboratory under Department of Energy Contract DE--AC02--06CH11357 and by the Extreme Science and Engineering Discovery Environment (XSEDE) at the Texas Advanced Computing Center under National Science Foundation contract TG--CHE130008.

\section{Appendix: Wavepacket Propagation}

\par The behavior of a wavepacket in the presence of a complex potential is readily determined through a real--time propagation scheme. Writing the packet wavefunction and complex potentials in terms of their real and imaginary parts, $\psi(x,t) = \text{Re}[\psi(x,t)] + i\text{Im}[\psi(x,t)]$ and  $V(x) = \text{Re}[V(x)] + i\text{Im}[V(x)]$ \cite{Visscher1991}, respectively, and substituting these into the single--particle Schr\"{o}dinger equation (with $\hbar = m = 1$)

\begin{equation}
i \frac{\partial \psi(x,t)}{\partial t} = -\frac{1}{2} \frac{\partial^2 \psi(x,t)}{\partial x^2} + \hat{V}(x) \psi(x,t),
\end{equation}

\noindent a coupled pair of equations for wavepacket evolution is obtained  after equating real and imaginary parts:

\begin{eqnarray}
\frac{\partial}{\partial t}\left[\text{Im}[\psi(x,t)]\right] &=& \frac{1}{2} \frac{\partial^2}{\partial x^2} \left[\text{Re}[\psi(x,t)]\right] + (\text{Im}[V(x)])(\text{Im}[\psi(x,t)]) \\
&&- (\text{Re}[V(x)])(\text{Re}[\psi(x,t)])\\
\frac{\partial}{\partial t}\left[\text{Re}[\psi(x,t)]\right] &=& -\frac{1}{2} \frac{\partial^2}{\partial x^2} \left[\text{Im}[\psi(x,t)]\right] + (\text{Im}[V(x)])(\text{Re}[\psi(x,t)]) \\
&&+ (\text{Re}[V(x)])(\text{Im}[\psi(x,t)]).
\end{eqnarray}

\noindent For the purposes of numerical evaluation, the derivatives are evaluated in a finite centered--difference approximation. Within such a scheme, the first derivative of the wavefunction is given by 

\begin{equation}
\frac{\partial}{\partial t} \psi(x,t) \approx \frac{\psi(x,t + \Delta t) - \psi(x,t-\Delta t)}{2 \Delta t}
\end{equation}

\noindent while the second derivative is

\begin{equation}
\frac{\partial^2}{\partial x^2} \psi(x,t) \approx \frac{\psi(x + \Delta x,t) - 2\psi(x,t) + \psi(x-\Delta x, t)}{(\Delta x)^2}.
\end{equation}

\noindent Given these approximations, the imaginary propagation equation becomes

\begin{multline}
[\text{Im} \, \psi(x,t+\Delta t)] = [\text{Im} \, \psi(x,t)] + s (\text{Re} \, [\psi(x + \Delta x, t)] - 2 [\text{Re} \, \psi(x,t)] +\\
 [\text{Re} \, \psi(x - \Delta x, t)]) + (\Delta t) ([\text{Im} \, V(x)][\text{Im}\, \psi(x,t)] - [\text{Re} \, V(x)][\text{Re}\, \psi(x,t)])
\end{multline}

\noindent where $s = \Delta t / 2(\Delta x)^2$ has been introduced as the parameter controlling integration.  The real term is evaluated similarly 

\begin{multline}
[\text{Re} \, \psi(x,t+ \Delta t)] = [\text{Re} \, \psi(x,t)] - s (\text{Im} \, [\psi(x + \Delta x, t )] -2 [\text{Im} \, \psi(x,t)] + \\ 
[\text{Im} \, \psi(x - \Delta x, t )]) +  (\Delta t)([\text{Re} \, V(x)] [\text{Im} \, \psi(x,t)] \\
 + [\text{Im}\, V(x)] [\text{Re} \, \psi(x,t)].
\end{multline} 
  
\noindent

\clearpage

\clearpage

\clearpage

\clearpage

\clearpage

\clearpage

\clearpage

\clearpage

\clearpage

\clearpage

\clearpage

\clearpage

\clearpage

\clearpage

\begin{table}[h]
\begin{center}
{\small
\begin{tabular}{|c||c|c|c|c|}
\hline
$V_\text{barrier}$ & $\mathcal{G}_\text{Ana}$  & $\mathcal{G}_\text{P1}$ & $\mathcal{G}_\text{P2-4}$ & $\mathcal{G}_\text{P1-9}$\\
\hline \hline
$2.50 \times 10^5$ & 0.314 & 0.315  & 0.315 & 0.315 \\
$5.00 \times 10^5$ & 0.306 & 0.306  & 0.307 & 0.309 \\
$7.50 \times 10^5$ & 0.262 & 0.281  & 0.282 & 0.288 \\
$10.0 \times 10^5$ & 0.232 & 0.219  & 0.212 & 0.233 \\
\hline
\end{tabular}
}
\end{center}
\caption{Comparison of analytically--determined conductances $\mathcal{G}_\text{Ana}$ with simulation--derived conductances for the first transmission event $\mathcal{G}_\text{P1}$, the mean of the subsequent three events $\mathcal{G}_\text{P2--4}$, and the collective mean of all simulated events $\mathcal{G}_\text{P1--9}$ for transmission through a barrier with $L_\text{barrier} = 0.1$ units at several barrier strengths $V_\text{barrier}$.}
\label{conduct}
\end{table}

\clearpage

\bibliography{manuscript}

\end{document}